\documentstyle[12pt]{article}                    
\newfont{\ffont}{msym10}                          
\newcommand{\beq}{\begin{equation}}               
\newcommand{\eeq}{\end{equation}}                 
\newcommand{\bqry}{\begin{eqnarray}}              
\newcommand{\eqry}{\end{eqnarray}}                
\newcommand{\bqryn}{\begin{eqnarray*}}            
\newcommand{\eqryn}{\end{eqnarray*}}              
\newcommand{\NL}{\nonumber \\}                    
\newcommand{\preprint}[1]{\begin{table}[t]        
            \begin{flushright}                    
            \begin{large}{#1}\end{large}          
            \end{flushright}                      
            \end{table}}                          
\newcommand{\PD}[2]                               
    {\frac{\partial^{#2}}{\partial #1^{#2}}}      
\begin{document}
\preprint{LA-UR-97-3795-rev}
\title{The Schwinger Nonet Mass and Sakurai Mass-Mixing Angle Formulae 
Reexamined}
\author{\\ L. Burakovsky\thanks{E-mail: BURAKOV@T5.LANL.GOV} \
and \ T. Goldman\thanks{E-mail: GOLDMAN@T5.LANL.GOV} \
\\  \\  Theoretical Division, MS B285 \\ Los Alamos National Laboratory \\ 
Los Alamos, NM 87545, USA \\}
\date{ }
\maketitle
\begin{abstract}
We study the origins of the inaccuracies of Schwinger's nonet mass, and the 
Sakurai mass-mixing angle, formulae for the pseudoscalar meson nonet, and 
suggest new versions of them, modified by the inclusion of the pseudoscalar 
decay constants. We use these new formulae to determine the pseudoscalar decay
constants and mixing angle. The results obtained, $f_8/f_\pi =
1.185\pm 0.040,$ $f_9/f_\pi =1.095\pm 0.020,$ $f_\eta /f_\pi =1.085\pm 
0.025,$ $f_{\eta ^{'}}/f_\pi =1.195\pm 0.035,$ $\theta =
(-21.4\pm 1.0)^o,$ are in excellent agreement with experiment.   
\end{abstract}
\bigskip
\bigskip
{\it Key words:} Schwinger's formula, Gell-Mann--Okubo, chiral Lagrangian, 
pseudoscalar 
\hspace*{0.85in}mesons \\

PACS: 12.39.Fe, 12.40.Yx, 12.90.+b, 14.40.Aq, 14.40.Cs
\newpage
\section{Introduction }
Schwinger's original nonet mass formula \cite{Sch} (here the symbol for the
meson stands either for its mass {\it or} mass squared),
\beq
(4K-3\eta -\pi )(3\eta ^{'}+\pi -4K)=8(K-\pi ) ^2,
\eeq
and the Sakurai mass-mixing angle formula \cite{Sakurai},
\beq
\tan ^2\theta =\frac{4K-3\eta -\pi }{3\eta ^{'}+\pi -4K},
\eeq
both relate the masses of the isovector $(\pi ),$ isodoublet $(K)$ and 
isoscalar mostly octet $(\eta )$ and mostly singlet $(\eta ^{'})$ states of a 
meson nonet, and the nonet mixing angle $(\theta ).$ We alert the reader that,
although we use notation suggestive of masses below, each formula is to be 
reprised in terms of mass or mass squared values in this introductory section.

The relations (1) and (2) are usually derived in the following way:
For a meson nonet, the isoscalar octet-singlet mass matrix,
\beq
{\cal M}=\left( 
\begin{array}{cc}
M_{88} & M_{89} \\
M_{89} & M_{99}
\end{array}
\right ) ,
\eeq 
is diagonalized by the masses of the physical $\eta $ and $\eta ^{'}$ states:
\bqry
{\cal M} & = & \left( 
\begin{array}{rr}
\cos \theta & \sin \theta \\
-\sin \theta & \cos \theta
\end{array}
\right) \left(
\begin{array}{cc}
\eta  & 0  \\
0 & \eta ^{'}
\end{array}
\right) \left(
\begin{array}{rr}
\cos \theta & -\sin \theta \\
\sin \theta & \cos \theta
\end{array}
\right)   \NL
 & = & \left(
\begin{array}{cc}
\cos ^2\theta \;\eta +\sin ^2\theta \;\eta ^{'} & \sin \theta \cos \theta \;\!
(\eta ^{'}-\eta ) \\
\sin \theta \cos \theta \;\!(\eta ^{'}-\eta )  & \sin ^2\theta \;\eta +\cos ^2
\theta \;\eta ^{'}
\end{array}
\right) ,
\eqry
where $\theta $ is the nonet mixing angle, which is determined by comparing 
the corresponding quadrants of the matrices (3) and (4), by any of the three
following relations: 
\beq
\tan ^2\theta =\frac{M_{88}-\eta }{\eta ^{'}-M_{88}},
\eeq
\beq
\tan ^2\theta =\frac{\eta ^{'}-M_{99}}{M_{99}-\eta },
\eeq
\beq
\sin 2\theta =\frac{2M_{89}}{\eta ^{'}-\eta }.
\eeq
It is easily seen that Eqs. (5) and (6) are identical, since, due to the trace
invariance of ${\cal M},$
\beq
\eta +\eta ^{'}=M_{88}+M_{99},
\eeq
and therefore, $M_{88}-\eta =\eta ^{'}-M_{99},$ and $\eta ^{'}-M_{88}=
M_{99}-\eta .$ Eliminating $\theta $ from (5),(7), or (6),(7), with the 
help of $\sin 2\theta =2\tan \theta /(1+\tan ^2\theta ),$ leads, respectively,
to
\beq
(\eta -M_{88})(M_{88}-\eta ^{'})=M_{89}^2,
\eeq
\beq
(M_{99}-\eta )(\eta ^{'}-M_{99})=M_{89}^2,
\eeq
which again are identical, through (8). 

We note that the ``ideal'' structure of a meson nonet,
\beq
(\eta =2K-\pi ,\;\eta ^{'}=\pi ),\;\;{\rm or}\;\;(\eta =\pi ,\;\eta ^{'}=2K-
\pi ),
\eeq
and the corresponding pure $q\bar{q}$ physical state valence flavor 
wavefunctions,
\beq
\omega _\eta =-s\bar{s},\;\;\omega _{\eta ^{'}}=\frac{u\bar{u}+d\bar{d}}{
\sqrt{2}}\equiv n\bar{n},\;\;{\rm or}\;\;\omega _\eta =n\bar{n},\;\;\omega _{
\eta ^{'}}=s\bar{s},
\eeq
given by the mixing
\bqry
\omega _\eta  & = & \omega _8\cos \theta -\omega _9\sin \theta , \NL
\omega _{\eta ^{'}} & = & \omega _8\sin \theta +\omega _9\cos \theta 
\eqry
with the ``ideal'' nonet mixing angle,
\beq
\theta =\arctan \frac{1}{\sqrt{2}}\cong 35.3^o,\;\;{\rm or}\;\;\theta =
-\arctan \sqrt{2}\cong -54.7^o,
\eeq
and the definitions
\beq
\omega _8=\frac{u\bar{u}+d\bar{d}-2s\bar{s}}{\sqrt{6}},\;\;\;\omega _9=
\frac{u\bar{u}+d\bar{d}+s\bar{s}}{\sqrt{3}},
\eeq
are the unique solution to Eqs. (9),(10) under the quark model inspired 
conditions
\beq
M_{88}=\frac{4K-\pi }{3},\;\;\;M_{99}=\frac{2K+\pi }{3},\;\;\;M_{89}=-\frac{
2\sqrt{2}}{3}(K-\pi ),
\eeq
where the first of the three relations in (16) is the standard Gell-Mann--Okubo
mass formula \cite{GMO}.

For all well established meson nonets, except the pseudoscalar (and, we expect,
scalar) one(s), both linear and quadratic versions of Eqs. (5)-(10) are in good
agreement with experiment. For example, for vector mesons, if one assumes the 
validity of the Gell-Mann--Okubo formula $\omega _8=(4K^\ast -\rho )/3,$ then
one obtains from Eq. (9) with the measured meson masses \cite{pdg}, $M_{89}=
-0.209\pm 0.001$ GeV$^2$ in the quadratic case, and $-0.113\pm 0.001$ GeV in 
the linear case. Note that this is entirely consistent with $-0.196\pm 0.005$ 
GeV$^2$ and $-0.118\pm 0.003$ GeV, respectively, which follow from the third 
element of (16). For tensor mesons, a similar comparison gives 
$-0.305\pm 0.020$ GeV$^2$ vs. $-0.287\pm 0.015$ GeV$^2,$ and $-0.105\pm 0.008$
GeV vs. $-0.104\pm 0.005$ GeV, respectively.  

However, for the pseudoscalar nonet, one obtains from Eq. (5) with meson masses
squared, $\theta \approx -11^o,$ in sharp disagreement with 
experiment, which favors the $\eta $-$\eta ^{'}$ mixing angle in the vicinity 
of $-20^o$ \cite{pdg,data,Abele}. Although using linear meson masses in Eq. (5)
does give $\theta \approx -24^o,$ in better agreement with 
data than its mass-squared counterpart, the value of $M_{89},$ as given by (9),
is now $-0.165\pm 0.004$ GeV, vs. $-2\sqrt{2}/3\;(K-\pi )=-0.338\pm 0.004$ GeV.
This emphasizes that neither Schwinger's nonet mass formula nor the mass-mixing
angle relations (including Sakurai's) (5)-(7) hold for the pseudoscalar nonet. 

It is well known, however, that the pseudoscalar (and, probably, scalar) mass 
spectrum does not follow the ``ideal'' structure, Eq. (16), since the mass of 
the pseudoscalar isoscalar singlet state is shifted up from its ``ideal'' 
value of $(2K+\pi )/3,$ presumably by the instanton-induced 't Hooft 
interaction \cite{tH} which breaks axial U(1) symmetry 
\cite{symbr,Dmitra,Dmitra2}. However, the use of $M_{99}=(2K+\pi )/3+A,$ $A\neq
0,$ in Eqs. (5)-(8) will again lead to Schwinger's formula (9), which does not
hold for the pseudoscalar mesons, as just demonstrated. [In fact, the 
structure of this formula does not depend at all on $M_{99},$ as seen in (9).]
Therefore, instanton, as well as any other effects which may shift the mass 
of the pseudoscalar isoscalar singlet state, cannot constitute the explanation
of the failure of Schwinger's quartic mass and the Sakurai mass-mixing angle 
formulae for the pseudoscalar nonet. We believe, however, that the following
analysis can resolve this problem.

\section{Pseudoscalar meson mass squared matrix}
It is known that the observed mass splitting among the pseudoscalar nonet may 
be induced (in terms of the $1/N_c$ expansion) by the following symmetry 
breaking terms \cite{symbr},
\beq
L^{(0)}_m=\frac{{\bar f}^2}{4}\left( B\;{\rm Tr\;\!M}\left( U+U^{\dagger }
\right) +\frac{\varepsilon }{6N_c}\left[ \;\!{\rm Tr}\left( \ln U-\ln U^{
\dagger }\right) \right] ^2\;\right) ,
\eeq
with M being the quark mass matrix, 
\beq
{\rm M}={\rm diag}\;(m_u,\;m_d,\;m_s)=m_s\;{\rm diag}\;(x,y,1),\;\;\;x\equiv 
\frac{m_u}{m_s},\;y\equiv \frac{m_d}{m_s},
\eeq
$N_c$ the number of colors, $\bar{f}$ the pseudoscalar decay constant in the 
limit of exact nonet symmetry, and $B,\varepsilon ={\rm const.}$ This symmetry
breaking Lagrangian term is to be added to the U(3)$_L\times $U(3)$_R$ 
invariant non-linear Lagrangian 
\beq
L^{(0)}=\frac{{\bar f}^2}{4}\;{\rm Tr}\;\!\left( \partial _\mu U\partial ^\mu 
U^\dagger \right) ,
\eeq
with $$U=\exp (i\pi /\bar{f}) ,\;\;\;\pi \equiv \lambda _a\pi ^a,\;\;a=0,1,
\ldots ,8,$$ which incorporates the constraints of current algebra for the 
light pseudoscalars $\pi ^a$ \cite{Georgi}. 

As pointed out in ref. \cite{KM}, 
chiral corrections can be important, the kaon mass being half the typical 1 
GeV chiral symmetry breaking scale. Such large corrections are clearly 
required from the study of the octet-singlet mass squared matrix $M^2.$ In the
isospin limit $x=y$ one has \cite{DGH} (with $m_n\equiv (m_u+m_d)/2)$
\beq
M^2=B\left(
\begin{array}{cc}
\frac{2}{3}(2m_s+m_n) & \frac{2\sqrt{2}}{3}(m_n-m_s) \\
\frac{2\sqrt{2}}{3}(m_n-m_s) & \frac{2}{3}(m_s+2m_n)+\frac{\varepsilon }{BN_c}
\end{array}
\right) ,
\eeq
which, on the most naive level, through the schematic Gell-Mann--Oakes-Renner 
relations (to first order in chiral symmetry breaking) \cite{GOR},
\bqry
\pi ^2 & = & 2\;\!B\;m_n, \NL
K^2 & = & B\;(m_s+m_n), \NL
\eta ^2 & = & \frac{2}{3}\;\!B\;(2m_s+m_n),
\eqry
which we discuss in more detail below, reduces to
\beq
M^2=\frac{1}{3}\left(
\begin{array}{cc}
4K^2-\pi ^2 & -2\sqrt{2}\;(K^2-\pi ^2) \\
-2\sqrt{2}\;(K^2-\pi ^2) & 2K^2+\pi ^2+3\tilde{A} 
\end{array}
\right) ,\;\;\;\tilde{A}\equiv \frac{\varepsilon }{BN_c}.
\eeq

Also, the Nambu--Jona-Lasinio model with the instanton-induced 't Hooft 
interaction, initiated by Hatsuda and Kunihiro \cite{HK} and Bernard {\it et 
al.} \cite{Bern}, and then extensively studied by Dmitrasinovic 
\cite{Dmitra,Dmitra2}, provides shifts of both the pseudoscalar and scalar
isoscalar singlet masses by the same amount, but in opposite directions (viz., 
the pseudoscalar isoscalar singlet mass is increased, while the scalar one is
decreased). Thus, for the pseudoscalar mesons, rather than using the 
model-dependent $\tilde{A},$ as defined in (22), we introduce the quantity $A$
below, which may be considered as the sum of all possible contributions to the
shift of the isoscalar singlet mass (from instanton effects, $1/N_c$-expansion
diagrams, gluon annihilation diagrams, etc.). 

Here, however, we suggest that the actual form of the mass squared matrix for 
the pseudoscalar mesons (and the corresponding symmetry breaking terms in 
(17)) is as follows:
\beq
\bar{f}^2M^2=\frac{1}{3}\left(
\begin{array}{cc}
4f_K^2K^2-f_\pi ^2\pi ^2 & -2\sqrt{2}\;(f_K^2K^2-f_\pi ^2\pi ^2) \\
-2\sqrt{2}\;(f_K^2K^2-f_\pi ^2\pi ^2) & 2f_K^2K^2+f_\pi ^2\pi ^2+3f_9^2A 
\end{array}
\right) ,
\eeq
where $f$'s are the pseudoscalar decay constants defined below. 

Indeed, the form of such a mass squared matrix is determined by the form of
Gell-Mann--Okubo type relations among the masses of the isovector, isodoublet,
and isoscalar octet and singlet states (which in our case are Eqs. (26)-(28)
below), since this matrix must be equivalent to that of the form (3), which in
the case we are considering is
\beq
\bar{f}^2M^2=\left(
\begin{array}{cc}
f_8^2\eta _{88}^2 & f_8f_9\eta _{89}^2 \\
f_8f_9\eta _{89}^2 & f_9^2\eta _{99}^2
\end{array}
\right) ,
\eeq
and is diagonalized by the physical $\eta $ and $\eta ^{'}$ meson masses and 
decay constants:
\beq
\bar{f}^2M^2=\left(
\begin{array}{cc}
f_\eta ^2\eta ^2 & 0  \\
0 & f_{\eta ^{'}}^2\eta ^{'2}
\end{array}
\right) .
\eeq
The equivalence of the matrices (23) and (24) is guaranteed by the validity of
the following relations: 
\beq
f_8^2\eta _{88}^2=-\frac{1}{3}\left[ m_u\langle \bar{u}u\rangle +m_d\langle 
\bar{d}d\rangle +4m_s\langle \bar{s}s\rangle \right] =\frac{4f_K^2K^2-f_\pi ^2
\pi ^2}{3}, 
\eeq
\beq
f_8f_9\eta _{89}^2=-\frac{\sqrt{2}}{3}\left[ m_u\langle \bar{u}u\rangle +m_d
\langle \bar{d}d\rangle -2m_s\langle \bar{s}s\rangle \right] =\frac{2\sqrt{
2}}{3}\left( f_\pi ^2\pi ^2-f_K^2K^2\right) ,
\eeq
\beq
f_9^2\eta _{99}^2=f_9^2A-\frac{2}{3}\left[ m_u\langle \bar{u}u\rangle +m_d
\langle \bar{d}d\rangle +m_s\langle \bar{s}s\rangle \right] =f_9^2A\;+\;\frac{
2f_K^2K^2+f_\pi ^2\pi ^2}{3},
\eeq 
with 
\beq
K^2\equiv \frac{(K^\pm )^2+(K^0)^2}{2},
\eeq
as suggested by Dmitrasinovic \cite{Dmitra2}, on the basis of the (precise)
Gell-Mann--Oakes-Renner formulae which relate the pseudoscalar
masses and decay constants to the quark masses and condensates \cite{GOR}:
\bqry
f_\pi ^2\;\pi ^2 & = & -\left[ m_u\langle \bar{u}u\rangle +m_d\langle \bar{d}d
\rangle \right] , \\
f_K^2\left( K^\pm \right) ^2 & = & -\left[ m_u\langle \bar{u}u\rangle +m_s
\langle \bar{s}s\rangle \right] , \\
f_K^2\left( K^0\right) ^2 & = & -\left[ m_d\langle \bar{d}d\rangle +m_s\langle
\bar{s}s\rangle \right] .
\eqry
(We ignore Dashen's theorem violating effects \cite{Dash}, and only 
approximately take into account isospin violating effects via (29), as we are
not concerned here with accuracies better than 1\%.)
 
Note that in the limit of exact nonet symmetry,
\beq
f_\pi =f_K=f_{88}=f_{99}\equiv \bar{f},\;\;\;\langle \bar{u}u\rangle =\langle 
\bar{d}d\rangle =\langle \bar{s}s\rangle \equiv \langle \bar{q}q\rangle ,
\eeq
one has the mass squared matrix (20) with $B=-\langle \bar{q}q\rangle /\bar{
f}^2,$ which further reduces to (22). The real world, and the mass squared 
matrix (23) associated with it, however (as we shall see), corresponds to the 
situation when the first set of the relations (33) is broken, but the second 
one remains (approximately) satisfied, i.e., the amount of SU(3) flavor 
symmetry breaking in terms of the quark condensates is much smaller than that 
in terms of the pseudoscalar decay constants.  

\section{Modified Gell-Mann--Okubo mass formula}
Here, we shall only explicitly demonstrate the validity of the modified 
Gell-Mann--Okubo formula (26) (the remaining relations (27) and (28) may be 
checked in a similar way).

First, this relation, as well as the mass squared matrix (23),(24), may be 
obtained through relating the vacuum expectation values of the equal time 
axial divergence-axial current commutators (which are ``sigma'' commutators in
Eq. (34),(35) below) to integrals over the pseudoscalar meson spectrum, as done
by Gensini \cite{Gen} following Gatto {\it et al.} \cite{Gatto}. For the
symmetry realized through a set of massless Goldstone pseudoscalar mesons,
only the pole terms survive in the first order of the symmetry breaking and
are expected to dominate over the continuum, which contributes only at the
second-order level. We therefore have the identities \cite{Gatto}
\beq
\langle \sigma _{ab}(0)\rangle =f_a f_b M^2_{ab}+\int d\mu \;\rho _{ab}(\mu ),
\eeq
which define the current quark mass contributions to the pseudoscalar masses.
These may be decomposed as
\beq
M^2_{ab}=\frac{\langle \sigma _{ab}(0)\rangle }{f_a f_b}+\left( M^2_{e.m.}
\right) _{ab}+\left( M^2_g\right) _{ab}+\left( M^2_{h.o.t.}\right) _{ab},
\eeq
where $M^2_{e.m.}$ is the long-range electromagnetic contribution, $M^2_g$ is
the gluon term which stands for both perturbative two-gluon annihilation and 
nonperturbative instanton effects. (Both are responsible for the shift of
the isoscalar singlet mass prior to its mixing with the isoscalar octet,
leading to the physical $\eta $ and $\eta ^{'}$ masses.) The latter must 
therefore act only on the singlet-singlet state matrix element:
\beq
\left( M^2_g\right) _{ab}=A\delta _{a9}\delta _{b9}.
\eeq
Finally, $M^2_{h.o.t.}$ stands for the contribution of the higher order terms.

Mass terms, which are the only explicit symmetry-breaking terms, have the
general form\footnote{We use the notations of ref. \cite{Gen}.}
\beq
L_m(x)=\varepsilon _0S_0(x)+\varepsilon _3S_3(x)+\ldots +\varepsilon _{N^2-1}
S_{N^2-1}(x),
\eeq
where $S_i=\bar{\psi }(x)\;\!\lambda _i/2\;\!\psi (x),$ $\lambda _0=(2/N)^{
1/2}$ {\bf I}, and $\lambda _{M^2-1}$ are the Gell-Mann-type matrices which 
constitute the SU(M) basis, $M\leq N.$ 
Restricting the indices to a subgroup SU($N^{'}),$ $N^{'}<N,$ does not
introduce higher mass quark fields anywhere but in the singlet-singlet part,
where they can be absorbed into the gluon term within the definition of 
higher-order terms given above. This restriction is meaningful only if the 
mixing of low mass $\bar{\psi }\psi $ pairs with higher mass ones is small 
enough. This is precisely what preserves the approximate SU(3) symmetry of
hadronic interactions, independent of the total number of flavors.

In terms of the current quark masses we also have
\bqry
\varepsilon _0 & = & \left( \frac{N}{2}\right) ^{-1/2}\sum _{i=1}^Nm_i, \\
\varepsilon _{M^2-1} & = & \left[ M(M-1)\right] ^{-1/2}\left( \sum _{i=1}^{M-1}
m_i\;-\;(M-1)\;m_M\right) .
\eqry
Assuming the SU(3)-invariant vacuum, fixing $M=3,$ and neglecting long-range
electromagnetic effects, one obtains in the lowest order of the symmetry 
breaking
\bqry
\left( \pi ^{\pm }\right) ^2 & = & \frac{\bar{f}^2}{f_\pi ^2}\;B\;(m_u+m_d), 
 \NL
\left( K^{\pm }\right) ^2 & = & \frac{\bar{f}^2}{f_K^2}\;B\;(m_u+m_s), \NL
\left( K^0\right) ^2 & = & \frac{\bar{f}^2}{f_K^2}\;B\;(m_d+m_s)
\eqry
in the charged sector, and
\bqry
\eta ^2_{33} & = & \frac{\bar{f}^2}{f_\pi ^2}\;B\;(m_u+m_d), \NL
\eta ^2_{38} & = & \frac{\bar{f}^2}{f_\pi f_8}\;B\;\frac{m_u-m_d}{\sqrt{3}},
 \NL
\eta ^2_{39} & = & \frac{\bar{f}^2}{f_\pi f_9}\;B\;\sqrt{\frac{2}{3}}\;(m_u-m_
d), \NL
\eta ^2_{88} & = & \frac{\bar{f}^2}{f_8^2}\;B\;\frac{4m_s+m_u+m_d}{3}, \NL
\eta ^2_{89} & = & -\;\frac{\bar{f}^2}{f_8 f_9}\;B\;\frac{\sqrt{2}}{3}\;(2m_s-
m_u-m_d), \NL
\eta ^2_{99} & = & \frac{\bar{f}^2}{f_9^2}\;B\;\frac{2}{3}\;(m_s+m_u+m_d)\;+\;A
\eqry
in the neutral $Y=I_3=0$ sector. Eliminating the quark
masses from the above two sets of relations, one obtains
\bqry
f_\pi ^2\eta ^2_{33} & = & f_\pi ^2\left( \pi ^{\pm }\right) ^2, \NL
f_8^2\eta ^2_{88} & = & \frac{1}{3}\;\left[ 2f_K^2\left( \left( K^{\pm }\right)
^2\;+\;2\left( K^0\right) ^2\right) \;-\;f_\pi ^2\left( \pi ^{\pm }\right) ^2
\right] , \NL
f_8f_9\eta ^2_{89} & = & -\;\frac{\sqrt{2}}{3}\;\left[ f_K^2\left( \left( K^{
\pm }\right) ^2\;+\;\left( K^0\right) ^2\right) \;-\;2f_\pi ^2\left( \pi ^{\pm
}\right) ^2\right] , \NL
f_9^2\eta ^2_{99} & = & \frac{1}{3}\;\left[ f_K^2\left( \left( K^{\pm }\right) 
^2\;+\;\left( K^0\right) ^2\right) \;+\;f_\pi ^2\left( \pi ^{\pm }\right) ^2
\right] \;+\;f_9^2A,
\eqry
which corresponds to Eqs. (26)-(28). Using the definition (20), one can also
obtain the mass squared matrix (23),(24) from Eqs. (40)-(42). (We have,
consistently, ignored small $\pi ^0-\eta $ and $\eta ^{'}$ $[\eta _{38},
\eta _{39}]$ mixing effects of electromagnetic and isospin breaking origin).

Independently, the relation (26) may be obtained from the following 
expressions for the pseudoscalar meson masses calculated by Li \cite{Li} in 
the chiral effective field theory of mesons $(\mu ,\Lambda =$ const):
\bqry
\pi ^2 & = & \frac{16N_c\;\!\mu ^3}{(4\pi )^2f_\pi ^2}\left( \ln \frac{\Lambda
^2}{\mu ^2}-\gamma +1\right) (m_u+m_d), \NL
\left( K^\pm \right) ^2 & = & \frac{16N_c\;\!\mu ^3}{(4\pi )^2f_K^2}\left( \ln
\frac{\Lambda ^2}{\mu ^2}-\gamma +1\right) (m_u+m_s), \NL
\left( K^0\right) ^2 & = & \frac{16N_c\;\!\mu ^3}{(4\pi )^2f_K^2}\left( \ln 
\frac{\Lambda ^2}{\mu ^2}-\gamma +1\right) (m_d+m_s), \NL
\eta _{88}^2 & = & \frac{16N_c\;\!\mu ^3}{(4\pi )^2f_{88}^2}\left( \ln \frac{
\Lambda ^2}{\mu ^2}-\gamma +1\right) \frac{m_u+m_d+4m_s}{3}.
\eqry 
Additionally, even at lowest order, $O(p^2),$ in generalized chiral 
perturbation theory, the inclusion of decay constants is required to form a 
relation among the octet pseudoscalar mesons \cite{KS}. Finally, this formula 
may be also obtained in standard chiral perturbation theory \cite{GL}, as 
follows.

Standard chiral perturbation theory leads to the following expressions for
the pseudoscalar meson masses and decay constants \cite{GL}:
\bqry
\pi ^2 & = & 2m_nB\left( 1+\mu _\pi -\frac{1}{3}\mu _8+2m_nK_3+K_4\right) ,
 \NL
K^2 & = & (m_n+m_s)B\left( 1+\frac{2}{3}\mu _8+(m_n+m_s)K_3+K_4\right) , \NL
\eta _{88}^2 & = & \frac{2}{3}(m_n+2m_s)B\left( 1+2\mu _K-\frac{4}{3}\mu _8
+\frac{2}{3}(m_n+2m_s)K_3+K_4\right) \NL
 & + & 2m_nB\left( -\mu _\pi+\frac{2}{3}\mu _K+\frac{1}{3}\mu _8\right) +
K_5, \NL
f_\pi & = & \bar{f}\left( 1-2\mu _\pi-\mu _K+2m_nK_6+K_7\right) , \NL
f_K & = & \bar{f}\left( 1-\frac{3}{4}\mu _\pi -\frac{3}{2}\mu _K-\frac{3}{4}
\mu _8+(m_n+m_s)K_6+K_7\right), \NL
f_8 & = & \bar{f}\left( 1-3\mu _K+\frac{2}{3}(m_n+2m_s)K_6+K_7\right) ,
\eqry
where $\mu $'s are chiral logarithms, and the constants $K_i$ are proper 
combinations of the low energy coupling constants $L_i.$ It then follows from 
these relations that the standard Gell-Mann--Okubo formula is broken in first
nonleading order \cite{GL},
\bqry
\triangle _{{\rm GMO}} & \equiv  & 4K^2-3\eta _{88}^2-\pi ^2\;=\;-2\left( 4K^2
\mu _K-3\eta _{88}^2\mu _8-\pi ^2\mu _\pi \right) +\ldots \NL
 & = & 4B\Big[ m_n(\mu _\pi +\mu _8-2\mu _K)+m_s(2\mu _8-2\mu _K)\Big] +\ldots
\;,
\eqry
where $\ldots $ stands for the higher order terms. 

However, the modified Gell-Mann--Okubo formula remains valid in this order, 
and is violated only by second order SU(3)-flavor breaking effects: 
\beq
\triangle ^{'}_{{\rm GMO}}\equiv \frac{1}{{\bar f}^2}\left( 4f_K^2K^2-3f_8^2
\eta _{88}^2-f_\pi ^2\pi ^2\right) =4B(m_s-m_n)\left( \mu _K+\frac{1}{2}\mu _8
-\frac{3}{2}\mu _\pi \right) +\ldots \;,
\eeq
since the second factor on the r.h.s. of (46) must vanish in the SU(3)-flavor
limit. Thus, the above analyses show, in an almost model independent way, that
the modified Gell-Mann--Okubo formula (26) is the only valid relation among the
octet pseudoscalar mesons,\footnote{A search for a relation of a more general 
form, $4f_K^aK^2=3f_8^a\eta _{88}^2+f_\pi ^a\pi ^2,$ which would hold in the 
first nonleading order of standard chiral perturbation theory, results in $a=
2.$} and therefore, the form of the mass squared matrix (23) is completely 
justified.\footnote{It has been suggested in the literature that the 
pseudoscalar decay constants should enter relations like (26)-(28) in the 
first rather than second power \cite{first}. As discussed above, such 
relations are expected to be less accurate than ours, according to chiral 
perturbation theory.}  
 
\section{The Schwinger and Sakurai formulae reexamined}
Starting with the mass squared matrix (23), the considerations which lead to
Eqs. (5)-(10) above, will now lead, through (26)-(28), to the following two 
relations,
\beq
\sin ^2\theta =\frac{4f_K^2K^2-3f_\eta ^2\eta ^2-f_\pi ^2\pi 
^2}{3f_{\eta ^{'}}^2\eta ^{'2}+f_\pi ^2\pi ^2-4f_K^2K^2},
\eeq
\beq
\left( 4f_K^2K^2-3f_\eta ^2\eta ^2-f_\pi ^2\pi ^2\right) \left( 3f_{\eta ^{
'}}^2\eta ^{'2}+f_\pi ^2\pi ^2-4f_K^2K^2\right) =8\left( f_K^2K^2-f_\pi ^2\pi 
^2\right) ^2,
\eeq
which we refer to as ``the Schwinger nonet mass, and Sakurai mass-mixing 
angle (respectively), formulae reexamined''. 

In contrast to $f_\pi $ and $f_K,$ the values of which are well established 
experimentally \cite{pdg}, 
\beq
\sqrt{2}f_K=159.8\pm 1.6\;{\rm MeV,}\;\;\;\sqrt{2}f_\pi=130.7\pm 0.3\;{\rm 
MeV},\;\;\;\frac{f_K}{f_\pi }=1.22\pm 0.01
\eeq
the values of $f_\eta ,$ $f_{\eta ^{'}}$ and $\theta $ are
known rather poorly. We now wish to calculate the values of $f_\eta ,$ $f_{
\eta ^{'}}$ and $\theta ,$ using the relations (47),(48), 
and compare the results with available experimental data. It is obvious that 
the two relations are not enough for determining the three unknowns. Note that
the additional relation, independent of (47),(48) (the trace condition for 
(23),(25)),
\beq
f_\eta ^2\eta ^2+f_{\eta ^{'}}^2\eta ^{'2}=2f_K^2K^2+f_9^2A,
\eeq
introduces an additional unknown, $A.$ We therefore develop another independent
relation among $f_\eta ,$ $f_{\eta ^{'}}$ and $\theta ,$ as follows. 

The light pseudoscalar decay constants are defined by the matrix elements
\beq
\langle 0|\bar{\psi }(0)\gamma ^\mu \gamma ^5\frac{\lambda ^j}{2}\psi (0)|P(p)
\rangle =i\delta ^{jP}f_Pp^\mu ,
\eeq
where $\psi =(u,d,s)$ is the fundamental representation of SU(3)$_f$, and $P=
(\pi ^0,\eta _{88},\eta _{99})$ means the corresponding SU(3)$_f$ indices 
3,8,9, thus picking out the diagonal $(j=3,8)$ SU(3)$_f$ Gell-Mann matrices  
$\lambda ^j,$ and $\lambda ^9\equiv \sqrt{2/3}$ {\bf I.} The neutral 
pseudoscalar wave functions, $P,$ may be expressed in terms of the quark basis
states $q\bar{q}$:
\beq
|P\rangle =\sum _q\frac{\lambda ^P_{q\bar{q}}}{\sqrt{2}}\;\!|q\bar{
q}\rangle \equiv \sum _qc^P_q|q\bar{q}\rangle ,\;\;\;q=u,d,s,
\eeq
where for $P=\pi ^0,$ $c^3_u=1/\sqrt{2}=\lambda ^3_{11}/\sqrt{2}=-\lambda ^3_{
22}/\sqrt{2}=-c^3_d,$ $c^3_s=0,$ for $P=\eta _{88},$ $c^8_u=c^8_d=1/\sqrt{
6}=\lambda ^8_{11}/\sqrt{2}=\lambda ^8_{22}/\sqrt{2},$ $c^8_s=-2/\sqrt{6}=
\lambda ^8_{33}/\sqrt{2},$ and for $P=\eta _{99},$ $c^9_u=c^9_d=c^9_s=1/\sqrt{
3}=(\lambda ^9/\sqrt{2})_{q\bar{q}}.$ 

The pseudoscalar decay constants defined in (51) can now be expressed as
\beq
f_P=\sum _q\frac{(\lambda ^P_{q\bar{q}})^2}{2}f_{q\bar{q}},
\eeq
where we have introduced the auxiliary decay constants $f_{q\bar{q}}$ defined 
as the decay constants of the $q\bar{q}$ pseudoscalar bound states having the
masses $M(q\bar{q}).$ In the isospin limit, $f_{u\bar{u}}=f_{d\bar{d}}=f_{u
\bar{d}}=f_{\pi ^0}=f_{\pi ^{+}}.$ Using this approximation, and evaluating 
the appropriate matrix elements leads to the following relations:
\bqry
f_\eta  & = & \left( \frac{\cos \theta -\sqrt{2}\sin \theta }{\sqrt{3}}\right)
^2f_\pi \;+\;\left( \frac{\sin \theta +\sqrt{2}\cos \theta }{\sqrt{3}}\right)
^2f_{s\bar{s}}, \\
f_{\eta ^{'}} & = & \left( \frac{\sin \theta +\sqrt{2}\cos \theta }{\sqrt{3}}
\right) ^2f_\pi \;+\;\left( \frac{\cos \theta -\sqrt{2}\sin \theta }{\sqrt{3}}
\right) ^2f_{s\bar{s}}.
\eqry  

Now we have four equations, (47),(48),(54),(55), which allow us to determine 
three unknowns, $f_\eta ,$ $f_{\eta ^{'}},$ $\theta ,$ as well
as the additional quantity introduced, namely, $f_{s\bar{s}}.$
The solution to these four equations is 
\bqry
\frac{f_\eta }{f_\pi } & = & 1.085\pm 0.025, \\
\frac{f_{\eta ^{'}}}{f_\pi } & = & 1.195\pm 0.035, \\
\frac{f_{s\bar{s}}}{f_\pi } & = & 1.280\pm 0.060, \\
\theta  & = & (-21.4\pm 1.0)^o.
\eqry
[The $\pi $ and $K$ electromagnetic mass differences, and the uncertainties in
the values of $f_\pi $ and $f_K,$ (see (49)), are taken as a measure of 
the uncertainties of the results.] 

Before comparing the solution obtained with
experiment, let us also calculate the values of $f_8$ and $f_9$ which are 
obtained from (54),(55) in the no-mixing case $(\theta =0):$
\bqry 
f_8 & = & \frac{1}{3}f_\pi \;+\;\frac{2}{3}f_{s\bar{s}}, \\
f_9 & = & \frac{2}{3}f_\pi \;+\;\frac{1}{3}f_{s\bar{s}}.
\eqry
Therefore, as follows from (58),(59),
\bqry
\frac{f_8}{f_\pi } & = & 1.185\pm 0.040, \\
\frac{f_9}{f_\pi } & = & 1.095\pm 0.020.
\eqry

The $\eta $-$\eta ^{'}$ mixing angle, as given in (59), is in agreement with
most of experimental data which concentrate around $-20^o$ 
\cite{pdg,data,Abele}. Also, the values for $f_8/f_\pi,$ $f_9/f_\pi $ and 
$\theta $ are consistent with those suggested in the literature, as we show in
Table I.

\begin{center}
\begin{tabular}{|c|c|c|c|c|} \hline
 Ref. & $f_8/f_\pi $ & $f_9/f_\pi $ & $\theta ,$ deg.  \\ 
\hline
 This work & $1.185\pm 0.040$ & $1.095\pm 0.020$ & $-21.4\pm 1.0$  \\ \hline
    [6]    & $1.11\pm 0.06$ & $1.10\pm 0.02$ & $-16.4\pm 1.2$   \\ \hline
  [25,26]  &     $1.25$     & $1.04\pm 0.04$ &   $-23\pm 3$     \\ \hline
    [27]   & $1.33\pm 0.02$ & $1.05\pm 0.04$ &   $-22\pm 3$     \\ \hline
    [28]   & $1.12\pm 0.14$ & $1.04\pm 0.08$ & $-18.9\pm 2.0$   \\ \hline
    [29]   & $1.38\pm 0.22$ & $1.06\pm 0.03$ & $-22.0\pm 3.3$   \\ \hline
    [30]   &     1.254      &     1.127      &    $-19.3$       \\ \hline
\end{tabular}
\end{center}
{\bf Table I.} Comparison of the values for $f_8/f_\pi,$ $f_9/f_\pi $ and
$\theta ,$ calculated in the paper, with the results of the papers referenced.
 \\
 
Note that (62),(63) are almost identical to (57),(56), respectively. This 
suggests that the quark content of the states must be the same. However, the 
relative phases, of course, cannot be. As we have suggested elsewhere 
\cite{content}, this conundrum may be resolved by identifying the 
wavefunctions in (15) with $\eta $ and $\eta ^{'}$ respectively, but with the 
signs of the $s\bar{s}$ terms of each reversed, viz., 
$$\eta \approx \frac{u\bar{u}+d\bar{d}-s\bar{s}}{\sqrt{3}},\;\;\;
\eta ^{'}\approx \frac{u\bar{u}+d\bar{d}+2s\bar{s}}{\sqrt{6}}.$$

\section{Comparison with data}
We now wish to compare the values obtained above for the ratios $f_8/f_\pi ,$ 
$f_9/f_\pi ,$ and for the $\eta $-$\eta ^{'}$ mixing angle with available 
experimental data. We shall first consider in more detail the well-known $\pi 
^0,\eta ,\eta ^{'}\rightarrow \gamma \gamma $ decays, for which experimental 
data are more complete than those for other processes involving light neutral 
pseudoscalar mesons, and then briefly mention the $\eta,\eta ^{'}\rightarrow
\pi ^{+}\pi ^{-}\gamma ,$ and $J/\psi \rightarrow \eta \gamma ,\eta ^{
'}\gamma $ decays.

\subsection{$P^0\rightarrow \gamma \gamma $ decays}
In the case of $\pi ^0\rightarrow \gamma \gamma ,$ the anomalous 
Wess-Zumino-Witten chiral lagrangian predicts \cite{WZW}
\beq
A_{\pi ^0\rightarrow \gamma \gamma }=\frac{\alpha N_c}{3\pi \bar{f}}\;\epsilon
^{\mu \nu \alpha \beta }\epsilon _{1\mu }\epsilon _{2\nu }k_{1\alpha }k_{2
\beta },
\eeq
where $\alpha =1/137.036$ is the fine structure constant, and $\epsilon _i,$ 
$k_i$ are the polarization and momenta of the outgoing photons. A leading-log 
calculation of the chiral corrections reveals that the dominant effect is 
simply to replace $\bar{f}$ by the physical value $f_\pi $ \cite{Hol}. The 
resulting amplitude is guaranteed by general theorems to remain unchanged in 
higher chiral orders \cite{AB}. One then finds that the predicted 
amplitude, as extracted from Eq. (71) below, with the experimentally measured
width \cite{pdg} $\Gamma (\pi ^0\rightarrow \gamma \gamma )=(7.7\pm 0.6)$ eV, 
\beq
F_{\pi \gamma \gamma }(0)=\frac{\alpha N_c}{3\pi f_\pi }=0.025\;{\rm GeV}^{-1},
\eeq
is in excellent agreement with experiment \cite{pdg}:
\beq
F_{\pi \gamma \gamma }(0)=(0.025\pm 0.001)\;{\rm GeV}^{-1},
\eeq
thus providing the confidence that one may analyze the corresponding 
$\eta ,\eta ^{'}$ decays with a similar precision.

In the case of the $\eta ,\eta ^{'}\rightarrow \gamma \gamma $ decays, one 
should include both the $\eta $-$\eta ^{'}$ mixing and the renormalization of 
the octet-singlet couplings, which leads to the predicted amplitudes
\bqry
F_{\eta \gamma \gamma }(0) & = & \frac{\alpha N_c}{3\sqrt{3}\pi f_\pi }\left(
\frac{f_\pi }{f_8}\cos \theta -2\sqrt{2}\frac{f_\pi }{f_9}\sin \theta \right) 
, \\
F_{\eta ^{'}\gamma \gamma }(0) & = & \frac{\alpha N_c}{3\sqrt{3}\pi f_\pi }
\left( \frac{f_\pi }{f_8}\sin \theta +2\sqrt{2}\frac{f_\pi }{f_9}\cos \theta 
\right) .
\eqry
The values of these amplitudes, as extracted from data, are \cite{VH}
\bqry
F_{\eta \gamma \gamma }(0) & = & 0.024\pm 0.001\;{\rm GeV}^{-1}, \NL
F_{\eta ^{'}\gamma \gamma }(0) & = & 0.031\pm 0.001\;{\rm GeV}^{-1}.
\eqry
Calculation with the help of Eqs. (40),(49),(54),(55),(58),(59) yields
\bqry
F_{\eta \gamma \gamma }(0) & = & 0.025\pm 0.001\;{\rm GeV}^{-1}, \NL
F_{\eta ^{'}\gamma \gamma }(0) & = & 0.030\pm 0.001\;{\rm GeV}^{-1},
\eqry
in excellent agreement with (69). Note that one can similarly compare the 
$\eta ,\eta ^{'}\rightarrow \gamma \gamma $ widths, as given by the relation
(see, e.g., ref. [13])
\beq
\Gamma (P^0\rightarrow \gamma \gamma )=\frac{F^2_{P^0\gamma \gamma}(0)M^2(P^
0)}{64\pi },
\eeq 
with $F_{P^0\gamma \gamma }(0)$ defined in (65),(67),(68), with those
measured. Such a comparison gives (in keV): $0.51\pm 0.05$ vs.
\cite{pdg} $0.46\pm 0.04$ for $\eta \rightarrow \gamma \gamma ,$ and $4.04\pm
0.27$ vs. \cite{pdg} $4.26\pm 0.19$ for $\eta ^{'}\rightarrow \gamma \gamma .$ 

Also, one can compare the values for the $\eta $-$\eta ^{'}$ 
mixing-independent $R$-ratio, given by (using (65),(67),(68),(71))
$$R\equiv \left[ \frac{\Gamma (\eta \rightarrow \gamma \gamma )}{\eta ^3}+
\frac{\Gamma (\eta ^{'}\rightarrow \gamma \gamma )}{\eta ^{'3}}\right] \frac{
\pi ^3}{\Gamma (\pi \rightarrow \gamma \gamma )}=\frac{1}{3}\left( \frac{f_
\pi ^2}{f_8^2}+8\frac{f_\pi ^2}{f_9^2}\right) .$$ 
With the values for $f_8/f_\pi $ and $f_9/f_\pi $ obtained above, our result 
is $R=2.25\pm 0.15,$ vs. $2.45\pm 0.35,$ as follows from using the 
experimentally measured masses and widths in the above expression.  

\subsection{$\eta,\eta ^{'}\rightarrow \pi ^{+}\pi ^{-}\gamma $ decays}
These, as well as the $P^0\rightarrow \gamma \gamma ,$ processes were 
extensively studied by Venugopal and Holstein \cite{VH} in chiral perturbation
theory. The analysis of experimental data for both of these processes done in 
ref. \cite{VH} yields
$$\frac{f_8}{f_\pi }=1.38\pm 0.22,\;\;\;\frac{f_9}{f_\pi }=1.06\pm 0.03,\;\;\;
\theta =(-22.0\pm 3.3)^o,$$ in good agreement with our Eqs. (59),(62),(63).

\subsection{$J/\psi \rightarrow \eta \gamma ,\eta ^{'}\gamma $ decays}
These processes were studied by Kisselev and Petrov \cite{KP}. The values of
$f_8/f_\pi ,$ $f_9/f_\pi $ and $\theta $ extracted in ref. 
\cite{KP} from the experimentally measured $P^0\rightarrow \gamma \gamma $ and
$J/\psi \rightarrow \eta \gamma ,\eta ^{'}\gamma $ widths, as given in the 
last column of Table I of ref. \cite{KP}, which corresponds to conventional 
mass-mixing angle relations, are
$$\frac{f_8}{f_\pi }=1.12\pm 0.14,\;\;\;\frac{f_9}{f_\pi }=1.04\pm 0.08,\;\;\;
\theta =(-18.9\pm 2.0)^o,$$ again in good agreement with our Eqs. 
(59),(62),(63). 

Thus, the three values of $f_8/f_\pi ,$ $f_9/f_\pi $ and $\theta $ agree with 
experiment (at least, as far as the processes considered above are concerned).
As to the remaining $f_\eta /f_\pi ,$ $f_{\eta ^{'}}/f_\pi $ ratios also 
calculated in the paper, the experimental values of them, as extracted from 
data by the CELLO \cite{CELLO} and TPC/2$\gamma $ \cite{TPC} collaborations, 
are, respectively,
\bqry
\frac{f_\eta }{f_\pi } & = & 1.12\pm 0.12, \\
\frac{f_{\eta ^{'}}}{f_\pi } & = & 1.06\pm 0.10,
\eqry
and
\bqry
\frac{f_\eta }{f_\pi } & = & 1.09\pm 0.10, \\
\frac{f_{\eta ^{'}}}{f_\pi } & = & 0.93\pm 0.09.
\eqry
While the value calculated for $f_\eta /f_\pi ,$ Eq. (56), clearly agrees with 
both experimental values (72) and (74), the value calculated for $f_{\eta ^{'}}
/f_\pi ,$ Eq. (57), only marginally agrees with (73), and disagrees with (75)
by almost 3 standard deviations.
 
To clarify this point, let us note that the values of $f_\eta /f_\pi $ and
$f_{\eta ^{'}}/f_\pi $ were extracted by both CELLO and TPC/2$\gamma $ from 
experimental data on the transition form-factors $T_{\eta (\eta ^{'})}(0,-Q^
2),$ assuming that the pole mass $\Lambda _{\eta (\eta ^{'})},$ which 
parametrizes their fits to the data, can be identified with $2\pi \sqrt{2}\;\!
f_{\eta (\eta ^{'})}.$ Then, these pole fits to the data are presumed to join 
smoothly, as $Q^2\rightarrow \infty ,$ to the perturbative QCD predictions for 
$T_{\eta (\eta ^{'})}(0,-Q^2)$ \cite{BL}, i.e., these fits would then agree 
with both the QCD asymptotic form $\sim 1/Q^2$ and its coefficient. However, 
the values of both $f_{\eta }$ and $f_{\eta ^{'}}$ quoted by the two groups, 
(in MeV) $(94.0\pm 7.1,$ $89.1\pm 4.9)$ \cite{CELLO}, and $(91.2\pm 5.7,$ 
$77.8\pm 4.9)$ \cite{TPC}, respectively, are all close to $M(\rho )/(2 \pi 
\sqrt{2})\approx 86.5$ MeV, thus indicating on possible connection with the 
vector meson dominance interpretation of $\Lambda _{\eta (\eta ^{'})}\approx 
M(\rho )$ in the range of $Q^2$ investigated. 

On the other hand, as remarked in ref. \cite{KK}, the model independent 
calculation by Gasser and Leutwyler \cite{GL}, testing the Goldberger-Treiman 
relations by Scadron \cite{Sca}, and the calculation by Burden {\it et al.} 
\cite{Burden}, all agree that both $f_\eta $ and $f_{\eta ^{'}}$ should be 
noticeably larger than $f_\pi .$ Our results $f_\eta \sim 1.1f_\pi ,$ $f_{
\eta ^{'}}\sim 1.2f_\pi $ are in agreement with this. 

It therefore seems that the extraction of the values of $f_\eta $ and $f_{\eta
^{'}}$ from the transition form-factors $T_{\eta (\eta ^{'})}(0,-Q^2)$ cannot 
be done accurately enough, at least in the range of $Q^2$ investigated so far.
That this may indeed be the case is indicated by the experimental value $f_{
\pi ^0}=84.1\pm 2.8$ MeV \cite{CELLO}, extracted by the same method, which is 
again close to $M(\rho )/2\pi \sqrt{2}.$ The central value of this $f_{\pi ^
0},$ 84.1 MeV, is $\sim 10$\% below the well established value given in Eq. 
(49), $92.4\pm 0.2$ MeV. Such a large discrepancy cannot be explained by, 
e.g., small isospin violation, indicating therefore that the extracted values 
for both $f_\eta $ and $f_{\eta ^{'}}$ may well have been underestimated too. 

\section{Concluding remarks}
As a lagniappe, we note that Eq. (50) may be combined with our extracted values
for $f_9,f_\eta,f_{\eta ^{'}}$ to obtain the value of $A:$
$$A=0.78\pm 0.12\;{\rm GeV}^2.$$
This is consistent with the usual value 0.73 GeV$^2$ determined by the trace
condition for (23),(25) without $f$'s \cite{content}.

Let us briefly summarize the findings of this work:

i) We have found that many theoretical approaches suggest that the more natural
object to study is the pseudoscalar mass squared matrix modified by the 
inclusion of the squared factors of the pseudoscalar decay constants.

ii) We have shown that this modified mass squared matrix leads to new 
Schwinger's quartic mass and the Sakurai mass-mixing angle relations for the 
pseudoscalar meson nonet.

iii) We have used these new relations for calculation of the pseudoscalar decay
constants and mixing angle. We have demonstrated that, except where questions
may be raised regarding the reliability of the extraction of the relevant 
quantities from direct experimental data, the results obtained are
in excellent agreement with available data.


\bigskip
\bigskip
\begin{thebibliography}{9}
\bibitem{Sch} J. Schwinger, Phys. Rev. Lett. {\bf 12} (1964) 237
\bibitem{Sakurai} J.J. Sakurai, {\it Currents and Mesons,} (University of
Chicago Press, Chicago, 1969)
\bibitem{GMO} S. Okubo, Prog. Theor. Phys. {\bf 27} (1962) 949, {\bf 28}
(1962) 24 \\ M. Gell-Mann and Y. Ne'eman, {\it The Eightfold Way,} (Benjamin,
NY, 1964)
\bibitem{pdg} Particle Data Group (R.M. Barnett {\it et al.}), Phys. Rev. D 
{\bf 54} (1996) 1
\bibitem{data} Crystal Barrel Collaboration (C. Amsler {\it et al.,}) Phys. 
Lett. B {\bf 294} (1992) 451 \\ P. Ball, J.-M. Frere and M. Tytgat, Phys. 
Lett. B {\bf 365} (1996) 367 \\ A. Bramon, R. Escribano and M.D. Scadron, 
Phys. Lett. B {\bf 403} (1997) 339; The $\eta $-$\eta ^{'}$ mixing angle
revisited, hep-ph/9711229
\bibitem{Abele} Crystal Barrel Coll. (A. Abele {\it et al.}), Phys. Lett. B
{\bf 402} (1997) 195
\bibitem{tH} G. 't Hooft, Phys. Rev. D {\bf 14} (1976) 3432, {\bf 18} (1978)
2199E
\bibitem{symbr} C. Rosenzweig, J. Schechter and C.G. Trahern, Phys. Rev. D 
{\bf 21} (1980) 3388 \\ P. Di Vecchia and G. Veneziano, Nucl. Phys. B 
{\bf 171} (1980) 253 \\ E. Witten, Ann. Phys. {\bf 128} (1980) 363
\bibitem{Dmitra} V. Dmitrasinovic, Phys. Rev. C {\bf 53} (1996) 1383 
\bibitem{Dmitra2} V. Dmitrasinovic, Phys. Rev. D {\bf 56} (1997) 247
\bibitem{Georgi} H. Georgi, {\it Weak Interactions and Modern Particle 
Theory,} (Benjamin, New York, 1984)
\bibitem{KM} D.B. Kaplan and A.V. Manohar, Phys. Rev. Lett. {\bf 56} (1986)
2004
\bibitem{DGH} J.F. Donoghue, E. Golowich and B.R. Holstein, {\it Dynamics of
the Standard Model,} (Cambridge University Press, 1996), Chapter X
\bibitem{GOR} M. Gell-Mann, R.J. Oakes and B. Renner, Phys. Rev. {\bf 175}
(1968) 2195
\bibitem{HK} T. Hatsuda and T. Kunihiro, Phys. Lett. B {\bf 206} (1988) 385,
erratum: {\it ibid.} {\bf 210} (1988) 278; Z. Phys. C {\bf 51} (1991) 49; 
Phys. Rep. {\bf 247} (1994) 221
\bibitem{Bern} V. Bernanrd, R.L. Jaffe and U.-G. Meissner, Nucl. Phys. B 
{\bf 308} (1988) 753
\bibitem{Dash} R.F. Dashen, Phys. Rev. {\bf 183} (1969) 1245
\bibitem{Gen} P.M. Gensini, J. Phys. G {\bf 7} (1981) 1315
\bibitem{Gatto} R. Gatto, G. Sartori and M. Tonin, Phys. Lett. B {\bf 28} 
(1968) 128
\bibitem{Li} Bing An Li, Phys. Rev. D {\bf 50} (1994) 2343
\bibitem{KS} M. Knecht and J. Stern, Generalized chiral perturbation theory,
hep-ph/9411253
\bibitem{GL} J. Gasser and H. Leutwyler, Nucl. Phys. B {\bf 250} (1985) 465 
\bibitem{first} G. Cicogna, F. Strocchi and R. Vergara Caffarelli, Phys. Rev.
D {\bf 6} (1972) 301 \\
T.N. Tiwari, C.V. Sastry and S. Sharma, Indian J. Pure Appl. Phys. {\bf 21}
(1983) 425
\bibitem{DHL} J.F. Donoghue, B.R. Holstein and Y.-C.R. Lin, Phys. Rev. Lett. 
{\bf 55} (1985) 2766
\bibitem{GK} F.J. Gilman and R. Kauffman, Phys. Rev. D {\bf 36} (1987) 2761
\bibitem{DGH2} Ref. [13], Chapter VII
\bibitem{KP} A.V. Kisselev and V.A. Petrov, Z. Phys. C {\bf 58} (1993) 595
\bibitem{VH} E.P. Venugopal and B.R. Holstein, Chiral anomaly and 
$\eta $-$\eta ^{'}$ mixing, hep-ph/9710382
\bibitem{CJ} H.-M. Choi and C.-R. Ji, Mixing angles and electromagnetic 
properties of ground state pseudoscalar and vector mesons in the light-cone
quark model, hep-ph/9711450
\bibitem{content} L. Burakovsky and T. Goldman, Gell-Mann--Okubo mass formula
revisited, hep-ph/9708498; Towards resolution of the scalar meson nonet enigma:
Gell-Mann--Okubo revisited, hep-ph/9709305, Nucl. Phys. A, {\it in press}
\bibitem{WZW} J. Wess and B. Zumino, Phys. Lett. B {\bf 37} (1971) 95 \\
E. Witten, Nucl. Phys. B {\bf 223} (1983) 422
\bibitem{Hol} B.R. Holstein, Phys. Lett. B {\bf 244} (1990) 83
\bibitem{AB} S. Adler and W.A. Bardeen, Phys. Rev. {\bf 182} (1969) 1517
\bibitem{CELLO} CELLO Coll. (H.-J. Behrend {\it et al.}), Z. Phys. C {\bf 49}
(1991) 401
\bibitem{TPC} TPC/2$\gamma $ Coll. (H. Aihara {\it et al.}), Phys. Rev. Lett.
{\bf 64} (1990) 172
\bibitem{BL} G.P. Lepage and S.J. Brodsky, Phys. Rev. D {\bf 22} (1980) 2157
 \\ S.J. Brodsky and G.P. Lepage, Phys. Rev. D {\bf 24} (1981) 1808
\bibitem{KK} D. Klabucar and D. Kekez, $\eta $ and $\eta ^{'}$ at the limits 
of applicability of a coupled Schwinger-Dyson and Bethe-Salpeter approach in
the ladder approximation, hep-ph/9710206
\bibitem{Sca} M.D. Scadron, Phys. Rev. D {\bf 29} (1984) 2076
\bibitem{Burden} C.J. Burden {\it et al.,} Phys. Rev. C {\bf 55} (1997) 2649
\end{thebibliography}
\end{document}